\documentclass[floatfix,twocolumn,aps,showpacs]{revtex4}
\def\beq{\begin{equation}}
\def\eeq{\end{equation}}
\def\bea{\begin{eqnarray}}
\def\eea{\end{eqnarray}}
\usepackage{epsfig}
\begin{document}

\title{Equation of State and Phases of Polarized Unitary Fermi Gas}

\author{S.~Y. Chang}
\affiliation{Department of Physics and Institute for Nuclear Theory,
             Box 351560, University of Washington, Seattle, WA 98195}

\begin{abstract}
 The equation of state of the partially polarized two component Fermi gas at zero temperature in the unitary
limit is computed by {\it ab initio} auxiliary field Monte Carlo method. From this, we obtain the critical ratio of the chemical potentials $\mu_\downarrow/\mu_\uparrow$ at the phase transitions. The value of $\mu_\downarrow/\mu_\uparrow$ at the transition between the fully paired superfluid and the partially polarized phases is $0.11$ while the critical value at the phase transition between the partially polarized phase and the fully polarized normal fluid is $-0.59$. We also determine the radial boundaries of the phase transitions of the Fermi gas in the harmonic trap as function of the total polarization. We find that beyond the critical polarization $0.65$, the fully paired superfluid core disappears in the trapped Fermi gas.
\end{abstract}
\maketitle

 Dilute fermion gases such as those of $^6$Li and $^{40}$K are quantum mechanical systems with controllable short range and strong interactions. They offer an ideal test bed for our knowledge of the quantum many-body physics.
The properties of these Fermi gases can be probed experimentally \cite{giorgini07}. 
The interaction is described by the dimensionless parameter $a_s k_F$ where $a_s$ is the s-wave scattering length
and $k_F $ is the Fermi momentum. In the weakly interacting (so-called BCS) regime where $1/a_sk_F << 0$ and up to the strongly interacting regime with $1/a_sk_F \approx 0$ (also called unitary regime), one or more non-trivial phases have been suggested \cite{wilczek03,fulde64,sarma63,sedrakian05} for the Fermi gases with spin imbalance. These polarized atomic gases hold resemblance to the case of magnetized superconductivity \cite{clogston62}. Here, instead of the external magnetic field, we assume unequal chemical potentials. The constraints given on the chemical potentials of the different fermion species suggest the existence of one or more intermediate polarized phases \cite{bulgac07}. However, these phases are hard to study theoretically since the mean field approaches are not quantitatively accurate, and the numerical techniques such as the Fixed Node Diffusion Monte Carlo (FN-DMC) method requires the knowledge of the physically motivated guiding functions in the first quantized form. 

\begin{figure}
\includegraphics[angle=0,width= 7.0cm]{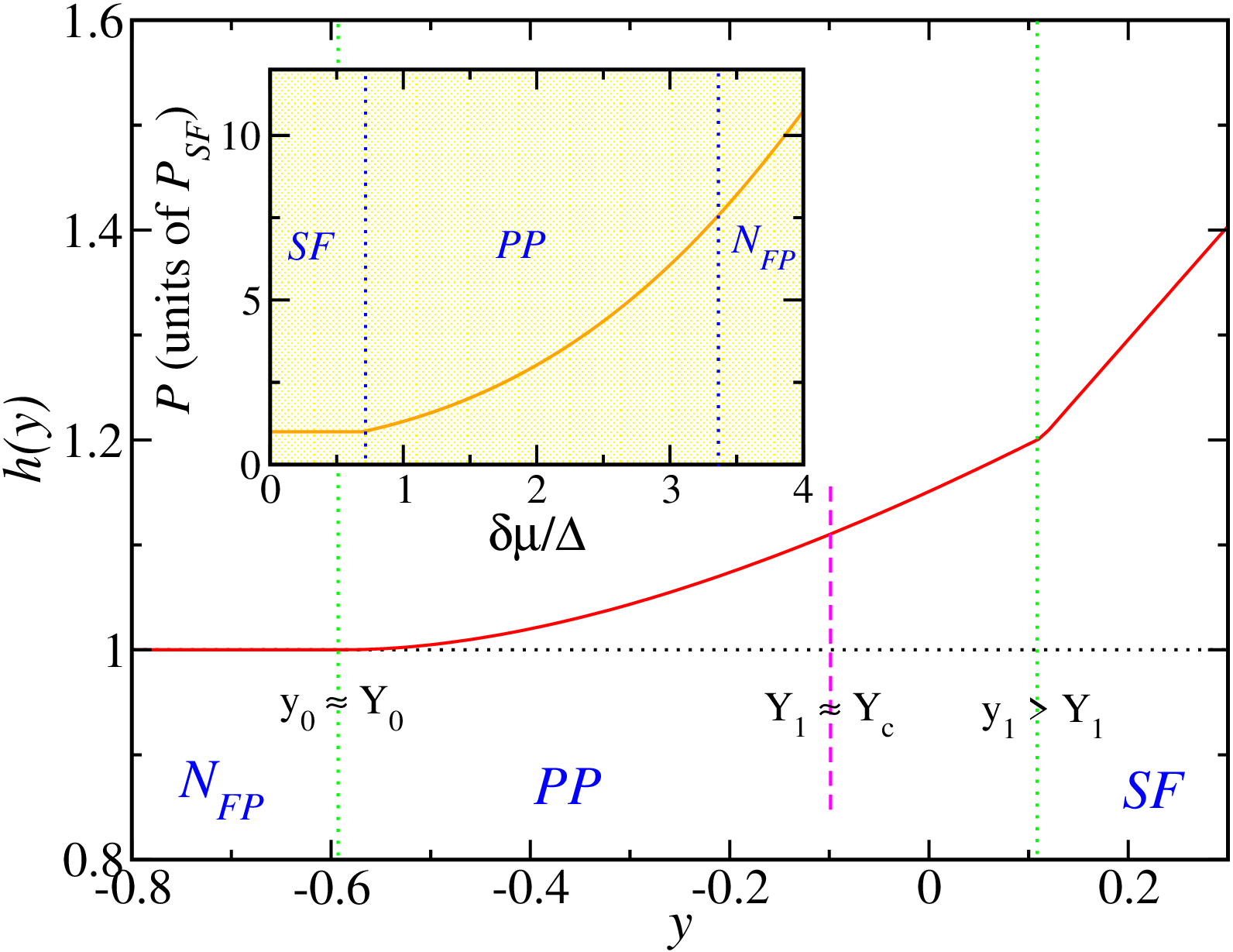}
\caption{(Color online)  We can clearly locate the values of $y_0 \approx Y_0$, $Y_1$ and $y_1$($> Y_1$) in this figure for $h(y)$ (Eq \ref{eqn_ep}). The discontinuity in $h'(y)$ (correspondingly in the densities) at $y_1$ is due to the first order nature of the phase transition between $SF$ and $PP$. For $y< y_0$, we have $N_{FP}$ phase, for $y_0 < y < y_1$ the $PP$ phase and for $y > y_1$ fully paired $SF$ phase. Here we assume $y_0 \approx Y_0$.
 The inset figure corresponds to the pressure in units of the superfluid pressure $P_{SF}$ as a function of $\delta \mu/\Delta$ (we take $\Delta/E_{FG} = 0.84$ \cite{carlson06}). Here we set $(\mu_\uparrow + \mu_\downarrow)/2$ constant. The {\it plateau} in the pressure for small values of $\delta \mu/\Delta$ is due to the existence of the pairing gap. It is the projection of the straight segment of $h(y)$ at $y > y_1$.} 
\label{fig_h}
\end{figure}

The system we consider in this article is that of the idealized Fermi gas consisting
of two($\uparrow$,$\downarrow$-spin) species with equal masses. We assume control on each one
of the chemical potentials ($\mu_\uparrow$,$\mu_\downarrow$) and the physically measurable quantities are the densities $n_\uparrow$ and $n_\downarrow$ ($0 \le n_\downarrow \le n_\uparrow$). In the spin symmetric phase, the existence of the gap is manifest in the fact that the densities are not sensitive to a small difference of the chemical potentials $\delta\mu \equiv (\mu_\uparrow - \mu_\downarrow)/2$. The superfluid phase imposes the constraint $\delta\mu \leq  \Delta$ \cite{cohen05}. Here $\Delta$ is the usual superfluid pairing gap of the symmetric system. This condition gives the lower bound on the critical $y \equiv \mu_\downarrow/\mu_\uparrow$ defined as $Y_1$ \cite{bulgac07}.
We use a capitalized notation $Y_x$ to indicate the upper or lower bounds while the lower case notation $y_x$ corresponds to the actual critical value at the phase transition. At a specific value of $y_1 \ge Y_1$ (or correspondingly at a critical $\delta\mu$), the fully paired superfluid ($SF$) undergoes phase transition into the partially polarized phase ($PP$, see Fig \ref{fig_h}) and the densities become unequal ($n_\downarrow < n_\uparrow$). Recently, two possibilities were considered. In the first, the $SF$ phase could transition into the polarized normal phase going through a phase separated mixture of the superfluid and the partially polarized normal fluid \cite{lobo06,bulgac07}. This transition that is assumed to be of the first order, is characteristic of the weakly interacting regime and also of the unitary limit. Another possibility is that the $SF$ phase undergoes the second order phase transition and becomes a homogeneous polarized superfluid that accommodates the excess of one species. This was suggested for small polarizations in the unitary limit by Carlson {\it et al.} \cite{carlson06}. This phase is alternatively called  gapless or polarized superfluid phase ($SF_p$). The recent work by Pilati {\it et al.} \cite{pilati07} considers both possibilities and suggests that the gapless homogeneous phase may occur in the $1/a_s k_F>0$ regime for moderate polarization. In the limit of the complete polarization, the system is in the normal fully polarized($N_{FP}$) phase with  $N_\uparrow$ spin up particles and $\mu_\uparrow = \frac{\hbar^2 (6 \pi^2n_\uparrow)^{2/3} }{2m} > 0$. This system is also insensitive to changes of $\delta \mu$ as long as $\mu_\downarrow <<0$. When $\mu_\downarrow \ge$ the energy difference between $N_\uparrow + 1_\downarrow$ and $N_\uparrow$ systems, the system phase transitions into the partially polarized($PP$) phase. This defines an upper bound for $y$ known as $Y_0$ \cite{bulgac07}. Several authors have shown \cite {lobo06,bulgac07,chevy06,chevy07,combescot07} that a simple variational solution of non-interacting $N_{FP}$ + interacting impurity gives an upper bound $Y_0$ reasonably close to the
actual threshold value $y_0$.

 In this letter, we construct the equation of state of the unitary Fermi gas connecting the limits 
$x\equiv\frac{n_\downarrow}{n_\uparrow}=0$ and $x=1$. Then we estimate the actual $y_1$ ($\ge Y_1$) and $y_0$ ($ \approx Y_0$) as a direct application of the knowledge of the equation of state. We also verify the consistency with the previously reported values of $Y_0$ and $Y_1$. We also present the equation of state in terms of the grand canonical potential (pressure) and the density profiles of the trapped gas. In order to do so, we implement the canonical ensemble auxiliary field Monte Carlo (AFMC) formalism at zero temperature. AFMC is usually formulated in the second quantized form. In principle, it does not depend on the particular choice of the basis. It can be applied to the finite temperature \cite{hirsch83,koonin93,bulgac06} as well as zero temperature \cite{koonin93,zhang97} Fermi systems. 
With the cost of introducing a set of additional integration variables, the time
propagator can be expressed in the basis set of one particle orbitals. Then, the multi-dimensional integrations over the additional auxiliary variables are carried out by the Monte Carlo method. The decomposition of the attractively interacting potential into negative eigenvalues avoids the essential sign problem associated with the complex time evolution matrix. However, for the imbalanced systems the sign problem appears for long enough time evolution for which we give a simple practical solution. We briefly outline the zero temperature canonical formalism with fixed particle numbers 
$N_\uparrow$ and $N_\downarrow$. We assume a zero range interaction of strength $g$ between the particles of different spin species. We have no exchange interaction and the Hamiltonian adopts the form
\bea
{\cal H} & = & \int d{\bf r} \left( \sum_{\sigma} -\hat\Psi^\dagger_\sigma({\bf r})\frac{\hbar^2 \nabla^2}{2m}\hat\Psi_\sigma({\bf r})\right) \ldots \nonumber \\
& &+ g\hat\Psi^\dagger_\uparrow({\bf r})\hat\Psi_\uparrow({\bf r})\hat\Psi^\dagger_\downarrow({\bf r})\hat\Psi_\downarrow({\bf r})
\eea	
where $\hat\Psi_\sigma({\bf r})$ and $\hat\Psi^\dagger_\sigma({\bf r})$ are the usual fermion field operators. We implement the solution of this Hamiltonian in a cubic volume of $N_l^3$ lattice sites. In this case, the constant $g$ is lattice renormalized coupling by the cutoff $k_c$ in the momentum space: $1/g = 1/g_b - 1/\Omega \sum_{|{\bf k}| =  0}^{k_c} 1/(2 \epsilon_{\bf k})$. Here, the bare coupling constant $g_b \equiv  4 \pi \hbar^2 a_s/m$, $\epsilon_{\bf k} \equiv \hbar^2 k^2/(2m)$, and $\Omega$ is the volume of the system.

The time evolution operator can be decomposed into a product of smaller steps 
$ e^{-\tau {\cal H}} = [e^{-\Delta \tau {\cal H}}]^M \approx [e^{-\Delta \tau {\cal T}/2} e^{-\Delta \tau {\cal V}}e^{-\Delta \tau {\cal T}/2}]^M$ where $\tau = M \Delta \tau$. The contributions of the one-body operators such as $ e^{-\Delta\tau {\cal T}/2}$ (with ${\cal T} \equiv - \sum_{\sigma} \hbar^2 \nabla^2/(2m) $) can be easily estimated in the one-body basis (that is, in the momentum space connected by Fourier transform). However, the interaction term ($e^{-\Delta \tau {\cal V}}$) needs to be treated before it becomes computable. 
Since the operator ${\hat n}_{\sigma}({\bf r}) \equiv \hat\Psi^\dagger_\sigma({\bf r})\hat\Psi_\sigma({\bf r})$ has eigenvalues 0 or 1, ${\hat n}^2_{\sigma}({\bf r}) = {\hat n}_{\sigma}({\bf r})$ and we can write \cite{hirsch83}
\beq
{\hat n}_{\uparrow}({\bf r}) {\hat n}_{\downarrow}({\bf r})= \frac{1}{2}({\hat n}_{\uparrow}({\bf r}) +{\hat n}_{\downarrow}({\bf r}))^2 -\frac{1}{2}({\hat n}_{\uparrow}({\bf r}) 
+{\hat n}_{\downarrow}({\bf r}))~.
\eeq
The last parenthesis is a sum of one-body operators. However, the first parenthesis is a square of one-body operators.
We introduce a set of one dimensional continuous variables at each lattice site $\bf r$ and use the Hubbard-Stratonovich transformation to get
\bea
& &e^{-\frac{\Delta \tau}{2} \sum\limits_{\bf r} g ({\hat n}_{\uparrow}({\bf r})+{\hat n}_{\downarrow}({\bf r}))^2 } \nonumber \\
 &=&  \prod\limits_{\bf r} \int\limits_{-\infty}^{\infty} dx({\bf r}) \frac{e^{-x({\bf r})^2/2}}{\sqrt{2\pi}}
e^{-x({\bf r}) \sqrt{-\Delta\tau g} ({\hat n}_{\uparrow}({\bf r})+{\hat n}_{\downarrow}({\bf r}))}~.
\label{eqn_hs}
\eea

\begin{figure}
\includegraphics[angle=0,width= 7.0cm]{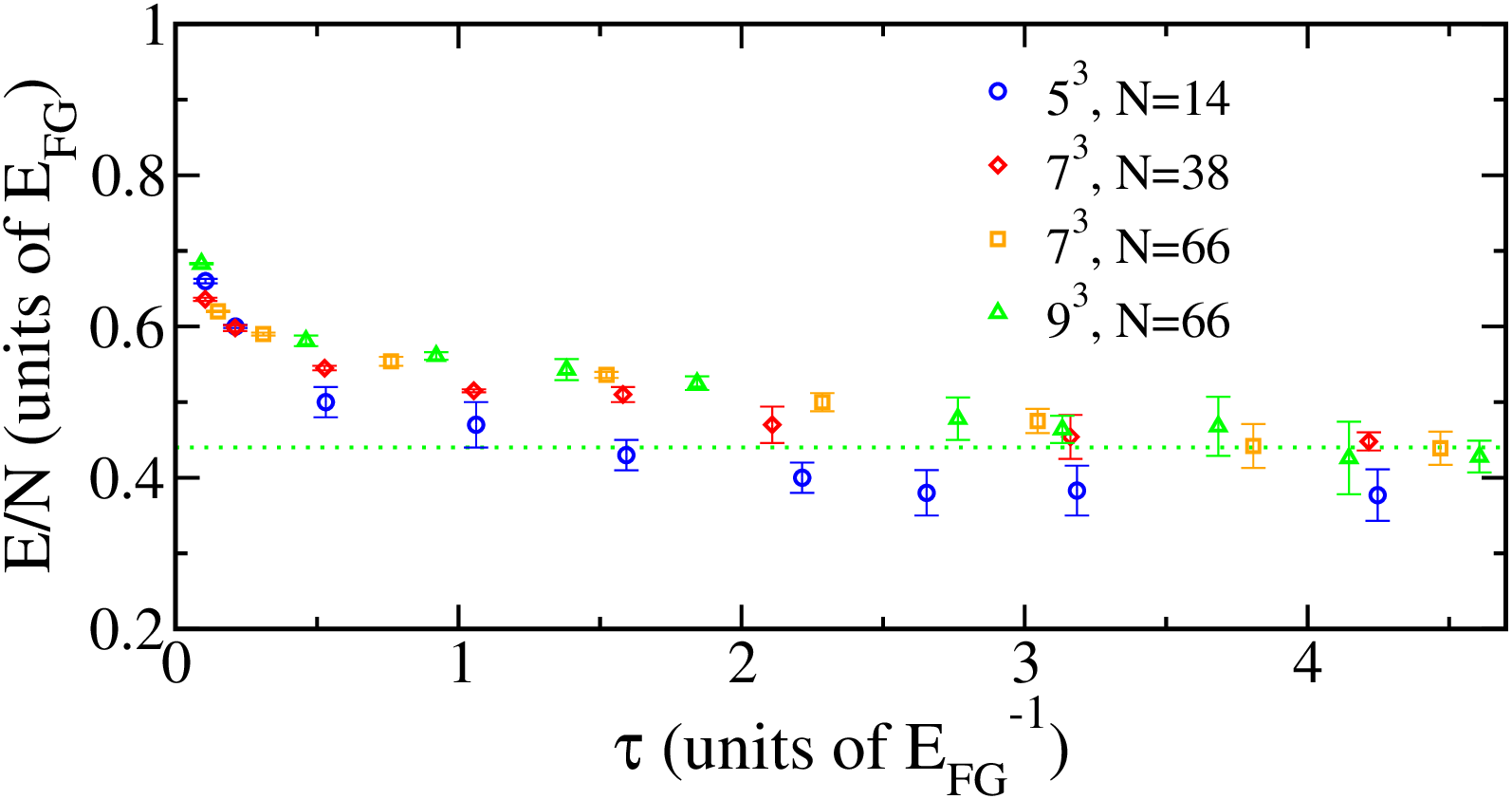}
\caption{(Color online) Imaginary time evolution for different lattice volumes $N_l^3$ and particle numbers $N = 2 N_\uparrow = 2 N_\downarrow$. We assume $\hbar = 1$. In general, the convergence is reached after $\tau E_{FG} \gtrsim 3$. We can see that for $N_l \ge 7$ and $N \ge 38$, the ground state energy per particle converges to $\sim 0.44$ in units of $ E_{FG}$.}
\label{fig_tevol}
\end{figure}

Then the time evolution operator can be expressed as
\beq
e^{-\tau {\cal H}} = \int {\cal D}[x] G[x] U[x]
\eeq
where the integration measure 
${\cal D}[x] = \prod_{{\bf r}, m} dx({\bf r},\tau_m) \frac{1}{\sqrt{2\pi}}$ ($0 \le m \le M$)
includes the auxiliary variables $\{x({\bf r},\tau_m)\}$ in the time slices $\{ \tau_m\}$. $G[x]$ is the Gaussian of the auxiliary variables: $G[x] = e^{-\frac{1}{2} \sum_{{\bf r},m} x^2({\bf r},\tau_m)}$.
Discrete auxiliary fields can also be used as shown in the references \cite{hirsch83,bulgac06}.
 The Gaussian factor $G[x]$ can be sampled directly leaving only the
integrand $U[x]$ that is used as the probability density for the Metropolis random walk.
From the stability argument we know that $E_0 >0$, thus
we have the bounds $\partial \langle {\cal H}(\tau)\rangle /\partial \tau <0$ and $\partial^2 \langle {\cal H}(\tau)\rangle /\partial^2 \tau >0$. The $\tau$ is pushed to the limit $\tau_0$ where the {\it plateau} with $\partial \langle {\cal H}(\tau)\rangle /\partial \tau \lesssim 0$ is reached (Fig \ref{fig_tevol}). 
Then we enter into the {\it sign problem} region where further evolution results in alternating signs
of the spin unmatched ($N_\uparrow \ne N_\downarrow$) fermion propagator. Instead of taking samples at various points of $\tau > \tau_0$, we perform separate runs with evolution up to $\tau_0$ to get the samples. We determine that the convergence to the ground state occurs after $\tau_0 E_{FG} \gtrsim 3$ ($\hbar$ set to 1). 

 In the zero temperature formalism, we assume an initial wave function that is not orthogonal to the ground state ${\bf \Psi}_t$. The ground state is projected out by taking $\lim\limits_{\tau \rightarrow \infty} e^{-\tau {\cal H}}{\bf \Psi}_t$. This is analogous to the DMC. However, here ${\bf \Psi}_t$ is a Slater determinant
represented by $N_b \times N$ matrix where $N_b$ is the size of the basis set and $N$ the number of particles. 
For the general quasiparticle creator $\hat c^\dagger_i = \sum_j D_{ji} \hat a^\dagger_j$ with $\hat a^\dagger_j = $ plane wave creation operator, $\{ D_{ji} \}$ are the elements of the matrix ${\bf \Psi}_t$ representing the
state $\prod_{i=1,N} c^\dagger_i |vac\rangle$. In our case, the initial state is constructed by completely filling the lowest $N_\uparrow$(and $N_\downarrow$) plane wave states with equal amplitude ($D_{ji} = \delta_{ji}$). 
The successive applications of the short time evolution operator yield a matrix of the same form and dimension. 
We can separately treat the different spin components of the wave function, ${\bf \Psi}_0   = \hat{U}_{\uparrow}(\tau) {\bf \Psi}_{t,\uparrow}  \bigotimes  \hat{U}_{\downarrow}(\tau) {\bf \Psi}_{t,\downarrow}$ and have the density operator in the momentum space 
\beq
 {\hat a}_{{\bf k}, \sigma}^\dagger {\hat a}_{{\bf k}', \sigma} = \left[{\bf \Psi}_{0,\sigma} ({\bf \Psi}_{t,\sigma}^T {\bf \Psi}_{0,\sigma})^{-1} {\bf  \Psi}_{t,\sigma}^T \right]_{{\bf k} {\bf k}'}~.
\eeq
Here $\hat{U}_{\sigma}(\tau)$ represents the  spin $\sigma$ time evolution operation and
${\bf \Psi}_{0,\sigma} \equiv \hat{U}_{\sigma}(\tau) {\bf \Psi}_{t,\sigma}$.
 The energy can be calculated as $\langle {\cal H} \rangle  = \sum\limits_{{\bf k},\sigma} \epsilon_{\bf k}  \langle {\hat a}_{{\bf k},\sigma}^\dagger {\hat a}_{{\bf k},\sigma}\rangle + g \sum\limits_{\bf r} \langle {\hat n}_{\uparrow}({\bf r}) {\hat n}_{\downarrow}({\bf r}) \rangle$. The sampling probability is $\langle \Psi_t| \Psi_0 \rangle$ = $det[{\bf \Psi}_t^T {\bf \Psi}_0]$ = $det[{\bf \Psi}_{t,\uparrow}^T {\bf \Psi}_{0,\uparrow}]\times det[{\bf \Psi}_{t,\downarrow}^T {\bf \Psi}_{0,\downarrow}]$. For $N_\uparrow \ne N_\downarrow$ and large enough $\tau$ this probability can be negative. Thus, we sample instead $|\langle \Psi_t| \Psi_0 \rangle|$ and the normalization becomes $\sum_{samples}\langle \Psi_t| \Psi_0 \rangle/|\langle \Psi_t| \Psi_0 \rangle|$. This sign problem may arise because ${\bf \Psi}_{t,\uparrow}$ and  ${\bf \Psi}_{t,\downarrow}$ can be different when $N_\uparrow \ne N_\downarrow$. We notice that for the case considered here where $N_\uparrow = 19$ and $\tau E_{FG} \sim 3$, the sign overlap of the determinants is still positively biased and the normalization non-zero. This is true even in the extreme polarization of $N_\uparrow + 1_\downarrow$ case. 
 
\begin{figure}
\includegraphics[angle=0,width= 7.0cm]{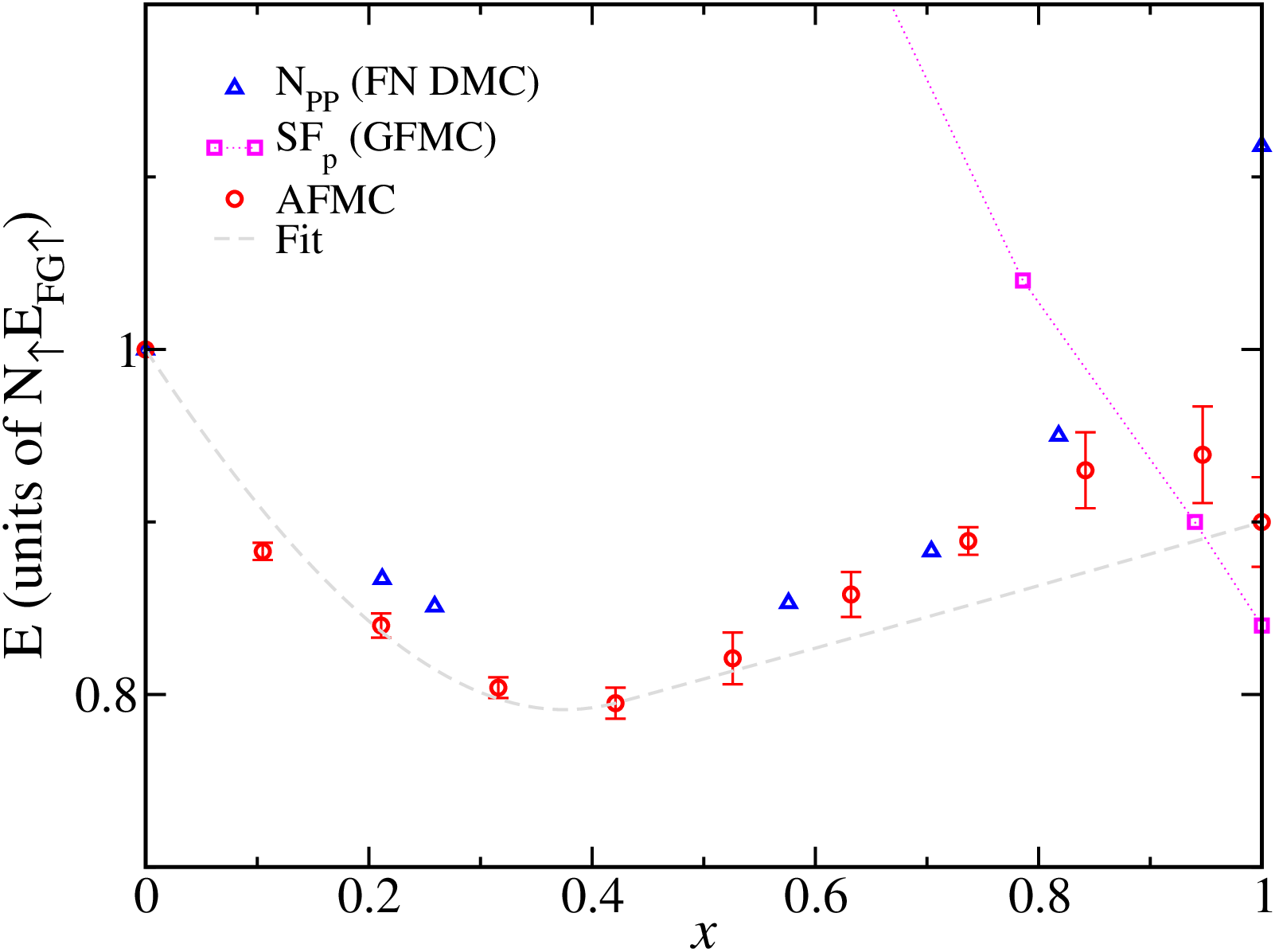}
\caption{(Color online) Equation of state of the unitary Fermi gas as function of the relative density $x$. We define the energy $E_{FG \uparrow} \equiv \frac{3}{5} \frac{\hbar^2 (6 \pi^2 n_\uparrow)^{2/3}}{2m}$. The AFMC results were obtained for $N_\uparrow = 19$ in a $7^3$ volume (circle). For the comparison purpose, we also plot the FN DMC results of $PP$ (triangle up, Ref \cite{lobo06}) and $SF_p$ (square, Ref \cite{carlson06}).The fit to the equation of state is consistent with the convexity of
the thermodynamic potential.}
\label{fig_results}
\end{figure}

\begin{figure}
\includegraphics[angle=0,width= 6.0cm]{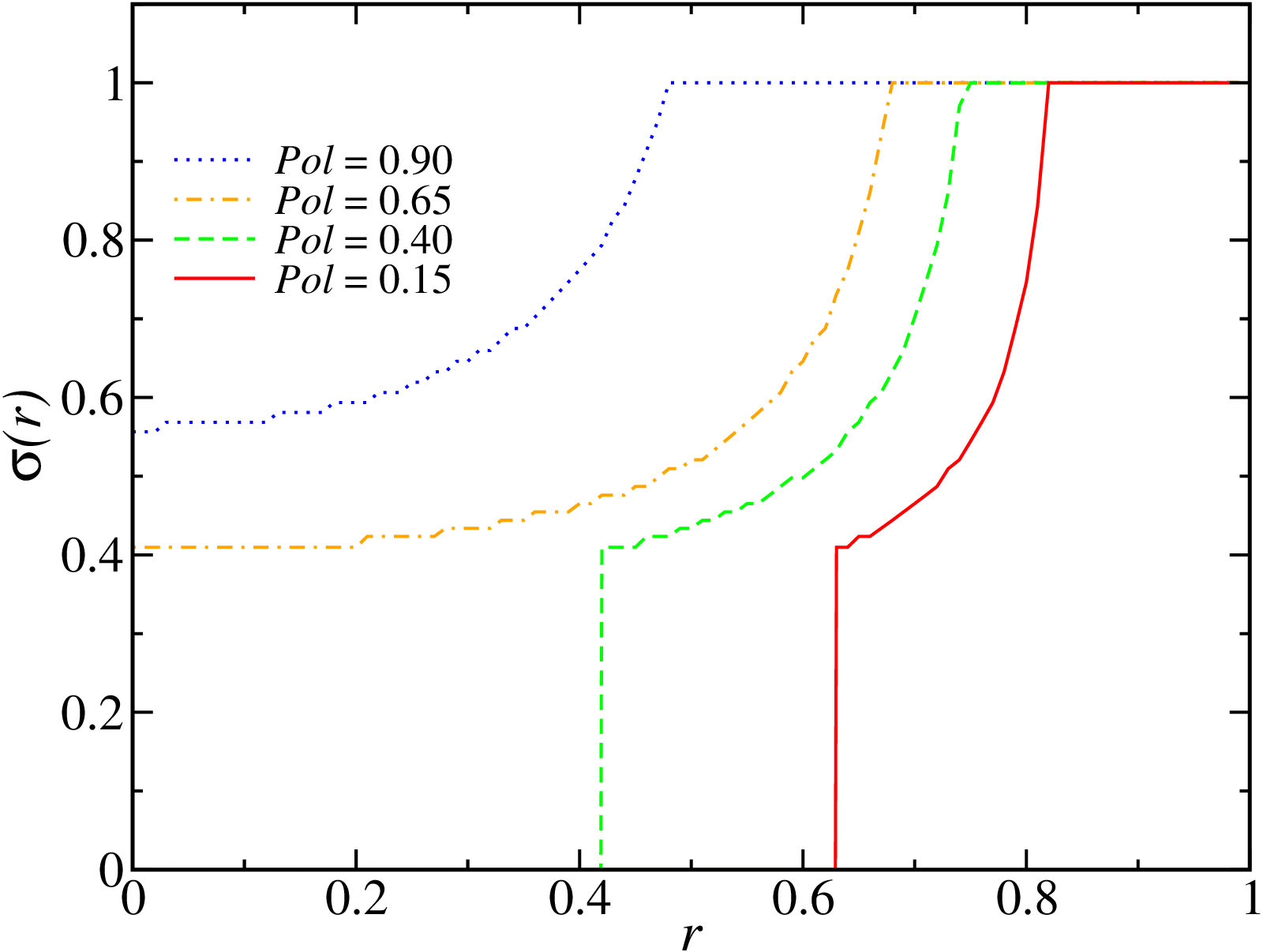}
\caption{(Color online) Local polarization $\sigma(r)$ at different total polarizations $Pol$. The radius $r$ is given in units of $r_{vac}$. $r_{vac}$ is defined as $\lambda_\uparrow = V(r_{vac})$. $Pol$ is controlled by the ratio of the global chemical potentials $\lambda_\downarrow/\lambda_\uparrow$. At the critical $Pol_c \approx 0.65$ we have the threshold where $r_1 \ge 0 $ appears (at smaller values of $Pol$). $r_1$ is
the boundary of the  fully paired core with $\sigma(r_1^-) = 0$ and $\sigma(r_1^+) >0$. At smaller polarization($Pol$), we can see the jump in the local density at $r_1$ corresponding to the first order phase transition $SF \leftrightarrow PP$. The radius $r_0$ equals the position where $\sigma(r_0) = 1$. This transition is of the second order and the $\sigma(r)$ behaves continuously.}
\label{fig_density}
\end{figure}

The possibility of the first order phase transition between $SF$ and $N_{FP}$ along $0\le x \le 1$ without any intermediate partially polarized phase($PP$) was discussed by Cohen \cite{cohen05}. However, from the constraints for $Y_0$ and $Y_1$ \cite{bulgac07} that establishes a strict inequality $Y_0 < Y_1$ this is not a likely scenario. Hence, we assume the existence of $PP$ phase and the  first order phase transition $SF \leftrightarrow PP$.
Then, we try to identify qualitatively different regions of $E(x)$. We identify that $x_c \approx 0.42$ separates those regions (see Fig \ref{fig_results}). At $x\lesssim 1$, apparently there is a {\it hump} in the energy which we interpret as the consequence of the finite size of the system. It would be mostly due
to the finite size pairing gap that will scale as $\sim \Delta/N_\uparrow$. In the range of $x_c \le x \le 1$, we fit the equation of state (data points at $x=x_c$ and $1$) with the functional form $f(x)$ consistent with the convexity constraint and the Maxwell construction \cite{cohen05,bulgac07}. Thus we take the form $f(x) = (a + b x)^{5/3}$ (see Eq \ref{eqn_ep}). All the data points in $0 \le x \le x_c$ were best fitted by a polynomial function of third power.  At $x=1$, $E/(N E_{FG}) = 0.44(2)$(with $E_{FG}\equiv \frac{3}{5} \frac{\hbar^2 (3 \pi^2 (n_\uparrow+n_\downarrow))^{2/3}}{2m}$) in close match with the previously known values \cite{carlson03,carlson06}. This number is free of any sign errors and convergent at different system sizes
and filling factors we have considered (see Fig \ref{fig_tevol}). The one particle chemical potential measured as $(E(N_\uparrow,1_\downarrow) -E(N_\uparrow,0_\downarrow))/E_{FG\uparrow}$ gives $-0.99 (2)$  consistent with the variational calculations \cite{lobo06,chevy06,chevy07,bulgac07,combescot07} and somewhat higher than Prokof'ev and others' diagrammatic Monte Carlo estimate of -1.03 \cite{proko07}. Our value corresponds to $y_0 \approx Y_0 = -0.59$. On the other hand, the value of $y_1$ calculated by taking $\mu_\downarrow/\mu_\uparrow$ at $x \rightarrow 1$ is $y_1 \approx 0.11$, well above the lower bound $Y_1 \approx -0.1$ and $Y_c = (2\xi)^{3/5}-1 \approx -0.1$. $Y_c$ was defined in the references \cite{cohen05,chevy06b} as the critical value for $y$ where if $Y_0 = Y_c = Y_1$ the $PP$ phase would
disappear. Clearly, this is not the case according to our results. Experiments seem to give a rather
wide range of possible values for the quantity $\gamma \equiv (1-y_1)/(1-y_0) = (1-r_0^2)/(1-r_1^2)$
(see Fig \ref{fig_density} for the definitions of $r_0$ and $r_1$ and the Ref \cite{bulgac07} for that of $\gamma$). $\gamma = 0.70$ in Ref \cite{zwierlein06} and $\gamma = 0.56$ in Ref \cite{shin07}.
In these experiments the effects of the finite temperature and the expansion introduce additional corrections.
 From our results of $y_0$ and $y_1$ we get $\gamma \approx 0.56$.
At the $SF \leftrightarrow PP$ phase transition $\delta\mu/E_{FG} = \frac{5}{3} \xi (1-y_1)/(1+y_1) \approx  0.58$ where $\xi$ is  the energy per particle in the $SF$ phase in units of $E_{FG}$. Thus, the superfluid pairs start to break at $\delta\mu/\Delta \geq 0.70$ (using $\Delta/E_{FG} =0.84(4) $ from reference \cite{carlson06}) while the completely polarized phase is reached for $\delta\mu/\Delta \geq 3.37$ (see Fig \ref{fig_h}).  
This picture is consistent with our earlier assumption that along the $x$ direction regions of pure $PP$ and mixture $PP$ + $SF$ exist rather than $N_{FP}$ + $SF$ mixture in the whole partially polarized region $0 < x < 1$. 

The free energy density of the polarized Fermi gas can be written as a function of the partial densities 
${\mathcal E}(n_\uparrow,n_\downarrow)$, where $n_\uparrow$ remains fixed and $n_\downarrow$
changes from $0$ to $n_\uparrow$. The chemical potentials are given by $\mu_{\sigma} = \partial{\mathcal E}/\partial n_{\sigma}$. Alternatively, the pressure can be obtained by the Legendre transform
${\mathcal P}(\mu_\uparrow,\mu_\downarrow) = \mu_\uparrow n_\uparrow + \mu_\downarrow n_\downarrow -{\mathcal E}
(n_\uparrow,n_\downarrow)$ where $n_\sigma = \partial {\mathcal P}/\partial \mu_\sigma $. From the dimensional argument, we can write the energy and the pressure in terms of the dimensionless functions $f(x)$ and $h(x)$ \cite{cohen05,bulgac07}
\bea
{\mathcal E}(n_\uparrow,n_\downarrow) & = & (6\pi^2)^{2/3}\frac{\hbar^2}{2m} n_\uparrow^{5/3} f(x)~, \nonumber \\
{\mathcal P}(\mu_\uparrow,\mu_\downarrow) & = & \frac{1}{15\pi^2}\left[ \frac{2m}{\hbar^2}\right]^{3/2}[\mu_\uparrow h(y)]^{5/2}~.
\label{eqn_ep}
\eea
Directly from the Monte Carlo output, we can give the equation of state either in the ${\mathcal E}$ vs $n_\sigma$ (see Fig \ref{fig_results}) or $\mathcal P$ vs $\mu_\sigma$(see Fig \ref{fig_h}) format. The $h(y)$ vs $y$ figure (Fig \ref{fig_h}) shows the pure phases in terms of $y$. In this figure, we can easily locate the relevant values of $y$. The inset figure for the pressure shows discontinuity in the slope
at the point of $SF \leftrightarrow PP$ phase transition while the slope is continuous at the transition
$PP \leftrightarrow N_{FP}$.

 For the case of the trapped gas, the local density approximation (LDA) defines the local chemical potentials $\mu_\sigma(r) = \lambda_\sigma - V(r)$ with $\lambda_\sigma =$ global chemical potential and $V(r) = \frac{1}{2} m \omega^2 r^2$, the harmonic trapping potential. Using the equation of state and the thermodynamic relations, we can establish the local density for each spin species $n_\sigma(r)=n_\sigma(\mu_\uparrow(r),\mu_\downarrow(r))$ in terms of $h(y)$ and $h'(y)$. We can define the local polarization $\sigma(r) = (n_\uparrow(r) - n_\downarrow(r))/(n_\uparrow(r) + n_\downarrow(r))$ and the total polarization $Pol= (N_\uparrow-N_\downarrow)/(N_\uparrow+N_\downarrow)$ for the trapped Fermi gas. Here $N_\sigma$ is obtained by integrating the local density $n_\sigma(r)$ over the
trapped volume. These are directly measurable quantities in the experiments (see Fig \ref{fig_density} and the reference \cite{shin07}). In our result, we observe that the radial boundary of the first order phase transition appears at $Pol$ lower than the critical $Pol_c \approx 0.65$. $Pol_c$ is defined when the radius $r_1 \rightarrow 0$ (see the caption of Fig \ref{fig_density}). This is qualitatively different from the value given by the mean field method \cite{yi06} where at the unitary regime the $Pol_c$ is at $\sim 1$ and the superfluid core appears at any $Pol < 1$. Also, the $\gamma \approx 0.85$ from the same mean field work in larger discrepancy with the experiments. Our $Pol_c$ is somewhat lower than those given by the references \cite{lobo06} and \cite{shin07} ($Pol_c \approx 0.77$ and $0.75$ respectively). The differences can be attributed to the calculation methods and the experimental conditions.

 In conclusion, we implemented a fully {\it ab initio} method for calculating the equation of state of the unitary Fermi gas at zero temperature. The sign problem of the spin imbalanced systems makes the Monte Carlo integration somewhat inefficient but still possible for the $N_\uparrow$ and $\tau$ considered. The comparisons at the extremes of the $N_\uparrow = N_\downarrow$ and $N_\downarrow=1$ with the available literature produce good match. This gave us confidence that in $0<x<1$, our result is close to the accurate equation of state. We were also able to extract the actual $y_1$ as the limit of $y(x)$ at $x \rightarrow 1$. By using the thermodynamic relations and LDA we could draw the local densities of the trapped gas where we can locate the phase transition radii at different total polarizations ($Pol$). The behavior of the pressure in different quantum phases was also studied. This method is general and we could easily extend to regimes off the unitary.

The author acknowledges helpful discussions with M.~M. Forbes. Also insightful comments from S. Reddy, J. Carlson, A. Bulgac, and  M. Boninsegni were useful.This work was  supported by the U.S. Department of Energy under Grants DE-FG02-00ER41132 and DE-FC02-07ER41457. The code was run in the UW-INT visitor cluster and the NERSC Jacquard supercomputing cluster.

\end{document}